\documentclass[11 pt]{article}
\usepackage[centertags]{amsmath}
\usepackage{amsfonts}
\usepackage{amssymb}
\usepackage{amsthm}
\usepackage{amsrefs}
\usepackage{hyperref}
\usepackage{newlfont}
\usepackage{xy}
\xyoption{all}
\usepackage{color}
\usepackage{epsfig}
\usepackage{fancyhdr}
\usepackage{graphicx}
\usepackage{float}
\usepackage{lscape}
\usepackage{pdflscape}

\newtheorem{theorem}{Theorem}[section]

\newtheorem{conc}[theorem]{Conclusion}
\newtheorem{rej}[theorem]{Rejection Criteria}

\newcommand{\RR}{\mathbb{R}}

\newcommand{\C}{\mathbb{C}}

\newcommand{\e}{\mathrm{e}}

\newcommand{\p}{\partial}

\newcommand{\cf}{\bar{f}}

\newcommand{\si}{\sigma}

\newcommand{\om}{\omega}

\newcommand{\Up}{\Upsilon}
\newcommand{\ka}{\kappa}

\newcommand{\glap}[1]{\Box_{#1}}

\renewcommand{\Re}{\operatorname{Re}}

\newcommand{\pnth}[1]{\left( #1 \right)}
\newcommand{\norm}[1]{\left\| #1 \right\|}
\newcommand{\abs}[1]{\left| #1 \right|}

\newcommand{\brce}[1]{\left\{ #1 \right\}}
\newcommand{\brkt}[1]{\left[ #1 \right]}

\title{Modeling Wave Dark Matter in Dwarf Spheroidal Galaxies}
\author{Hubert L. Bray\footnote{Mathematics and Physics Departments, Duke University, Box 90320, Durham, NC 27708, USA, \mbox{bray@math.duke.edu}}, \ Alan R. Parry\footnote{Mathematics Department, Duke University, Box 90320, Durham, NC 27708, USA, \mbox{alrparry@math.duke.edu}}}
\date{\today}

\begin{document}

\maketitle

\begin{abstract}
  This paper studies a model of dark matter called wave dark matter (also known as scalar field dark matter and boson stars) which has recently also been motivated by a new geometric perspective by Bray \cite{Bray10}.  Wave dark matter describes dark matter as a scalar field which satisfies the Einstein-Klein-Gordon equations.  These equations rely on a fundamental constant $\Up$ (also known as the ``mass term'' of the Klein-Gordon equation).  In this work, we compare the wave dark matter model to observations to obtain a working value of $\Up$.

  Specifically, we compare the mass profiles of spherically symmetric static states of wave dark matter to the Burkert mass profiles that have been shown by Salucci et al.\ \cite{Salucci11} to predict well the velocity dispersion profiles of the eight classical dwarf spheroidal galaxies.  We show that a reasonable working value for the fundamental constant in the wave dark matter model is $\Up = 50 \text{ yr}^{-1}$.  We also show that under precise assumptions the value of $\Up$ can be bounded above by $1000 \text{ yr}^{-1}$.
\end{abstract}

\section{Introduction}

Ever since the first postulation of dark matter in the 1930's by Zwicky \cite{Zwicky33}, much evidence for the existence of dark matter has accumulated including the unexpected behavior in the rotation curves of spiral galaxies \cites{Begeman89,Bosma81}, the velocity dispersion profiles of dwarf spheroidal galaxies \cites{Walker07,Salucci11,Walker09,Walker10}, and gravitational lensing \cite{Dahle07}.  These and other observations support the idea that most of the matter in the universe is not baryonic, but is, in fact, some form of exotic dark matter and that almost all astronomical objects from the galactic scale and up contain a significant amount of this dark matter.  Describing this dark matter is currently one of the biggest open problems in astrophysics \cites{DMAW,Hooper08,Bertone05,Ostriker93,Trimble87,Binney98,Binney08}.

In the last two decades, there has been substantial progress on describing the distribution of dark matter on the galactic scale.  Navarro, Frenk, and White's popular model \cite{NFW96}, which resulted from detailed $N$-body simulations, has been shown to agree well with observations outside the centers of galaxies \cites{Humphrey06,Walker09}.  However, the Navarro-Frenk-White dark matter energy density profile also exhibits an infinite cusp at the origin, while observations favor a bounded value of the dark matter energy density at the centers of galaxies \cites{Blok10,Gentile04}.  This has prompted many astrophysicists to employ a cored profile, such as the Burkert profile \cite{Burkert95}, which ``cores'' out the infinite cusp, to model the dark matter energy density in a galaxy.  Resolving this ``core-cusp problem'' remains an important open problem in the study of dark matter.

One possible solution to this problem, and potentially other astrophysical problems, is the introduction of a new model for dark matter.  Recently, Bray has geometrically motivated the study of a scalar field satisfying the Einstein-Klein-Gordon equations as a viable dark matter candidate via constructing axioms for general relativity \cite{Bray10}.  The idea of using a scalar field to describe dark matter is not new.  In fact, such a model has been seriously considered as a candidate for dark matter for more than two decades and has been shown to be in agreement with many cosmological observations \cites{Bray10,Lee09,Matos09,MSBS,Seidel90,Seidel98,Bernal08,Sin94,Mielke03,Sharma08,Ji94,Lee92,Lee96,Guzman01}.  In most of these settings, these scalar fields are considered from a quantum mechanical motivation for the same Einstein-Klein-Gordon equations and go by the name scalar field dark matter or boson stars.  However, due to the fact that the Klein-Gordon equation is a wave-type partial differential equation, we prefer the name {\em wave dark matter}.  In this paper, we begin to test wave dark matter against observations at the galactic level.  In particular, we seek a working estimate of the fundamental constant in the wave dark matter model, $\Up$, to be used in future comparisons to data.  To do so, we will compare the simplest model defined by wave dark matter to models of dark matter that are already known to fit observations well.

\begin{figure}

    \begin{center}
        \includegraphics[width = 6 in]{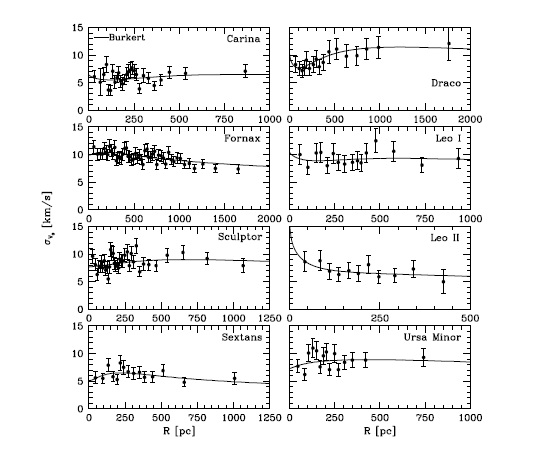}
    \end{center}

    \caption{Observed velocity dispersion profiles of the eight classical dwarf spheroidal galaxies are denoted by the points on each plot with its associated error bars.  The solid lines overlayed on these profiles are the best fit velocity dispersion profiles predicted by the Burkert mass profile.  This figure is directly reproduced from the paper by Salucci et al.\ \cite{Salucci11} and the reader is referred to their paper for a complete description of how these models were computed.}

    \label{vel-disp}

\end{figure}

Salucci et al.\ recently used the Burkert profile to model the dark matter energy density profiles of the eight classical dwarf spheroidal galaxies orbiting the Milky Way.  They found excellent agreement between the observed velocity dispersion profiles of these galaxies and those velocity dispersion profiles predicted by the Burkert profile \cite{Salucci11}.  This can be seen in Figure \ref{vel-disp}, which we have reproduced exactly as it appears in the paper by Salucci et al.  In what follows, we will show that a value of $\Up = 50 \text{ yr}^{-1}$ produces wave dark matter mass models that are qualitatively similar to the Burkert mass models found by Salucci et al.  We will also show that under precise assumptions, comparisons to these Burkert profiles can be used to bound the value of $\Up$ above by $1000 \text{ yr}^{-1}$.

\section{Burkert Mass Profiles}

\begin{figure}

  \begin{center}
    \includegraphics[height = 2.25 in, width = 3 in]{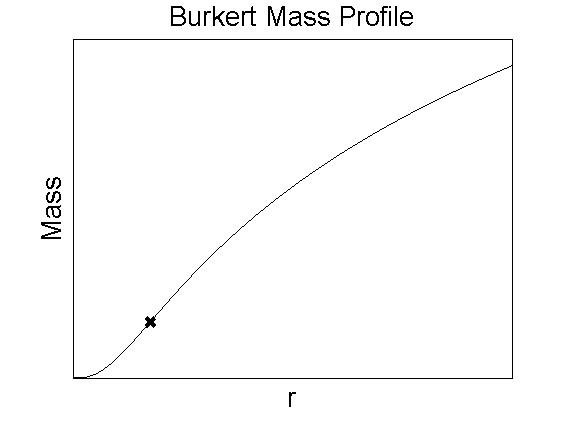}
  \end{center}

  \caption{Plot of a Burkert mass profile.  The inflection point is marked with an $\times$.}

  \label{BurMass}

\end{figure}

The Burkert energy density profile models the energy density of a spherically symmetric dark matter halo using the function
\begin{equation}\label{Burmod}
  \mu_{B}(r) = \frac{\rho_{0}r_{c}^{3}}{(r+r_{c})(r^{2}+r_{c}^{2})}
\end{equation}
where $\rho_{0}$ is the central density and $r_{c}$ is the core radius.  Integrating this function over the ball of radius $r$ with respect to the standard spherical volume form yields the Burkert mass profile as follows.
  \begin{align}
    \notag M_{B}(r) &= \int_{B_{r}(0)} \mu_{B}(s)\, dV_{\RR^{3}} \\
    \notag &= 4\pi \int_{0}^{r} s^{2}\mu_{B}(s)\, ds \displaybreak[0] \\
    \notag &= 4\pi \int_{0}^{r} \frac{s^{2}\rho_{0}r_{c}^{3}}{(s+r_{c})(s^{2}+r_{c}^{2})}\, ds \\
    \label{Bur-M} M_{B}(r) &= 2\pi \rho_{0}r_{c}^{3}\pnth{\ln\pnth{\frac{r+r_{c}}{r_{c}}} + \frac{1}{2}\ln\pnth{\frac{r^{2}+r_{c}^{2}}{r_{c}^{2}}} - \arctan\pnth{\frac{r}{r_{c}}}}
  \end{align}
A generic plot of a Burkert mass profile, $M_{B}(r)$, defined to be the dark matter mass in the ball of radius $r$, is shown in Figure \ref{BurMass}.  We make a few remarks about the behavior of this mass function.

Note that the behavior of the graph changes concavity at the inflection point $r = r_{ip}$, which we have marked on the plot in Figure \ref{BurMass} with an $\times$.  Recalling from equation (\ref{Bur-M}) the fact that $M_{B}(r)$ is the integral over the interval $[0,r]$ of the function $4\pi r^{2}\mu_{B}(r)$, we can compute this inflection point as follows.
\begin{equation}
  M_{B}'(r) = 4\pi r^{2}\mu_{B}(r) = \frac{4\pi\rho_{0}r_{c}^{3}r^{2}}{(r+r_{c})(r^{2}+r_{c}^{2})}
\end{equation}
Differentiating again yields
\begin{equation}
  M_{B}''(r) = \frac{-4\pi\rho_{0}r_{c}^{3}(r^{4} - r^{2}r_{c}^{2} - 2rr_{c}^{3})}{(r+r_{c})^{2}(r^{2}+r_{c}^{2})^{2}},
\end{equation}
which has two complex zeros and two real zeros.  The two real zeros are $r=0$ and
\begin{equation}\label{Bur_ip}
  r_{ip} = \pnth{\frac{3 + (27 + 3\sqrt{78})^{2/3}}{3(27 + 3\sqrt{78})^{1/3}}}r_{c} \approx 1.52r_{c},
\end{equation}
the latter being the inflection point of the mass model.

For $r \gg r_{ip}$, the plot grows logarithmically due to the fact that the $\arctan$ term in equation~(\ref{Bur-M}) approaches a constant value as $r \to \infty$.  To describe the behavior when $r \ll r_{ip}$, we note that the Taylor expansion of $M_{B}(r)$ centered at $r = 0$ is as follows,
\begin{equation}
  M_{B}(r) = \frac{4}{3}\pi \rho_{0}r^{3} + O(r^4).
\end{equation}
Thus for $r \ll r_{ip}$, $M_{B}(r)$ is dominated by an $r^{3}$ term making the initial behavior cubic.

\begin{figure}

  \begin{center}
    \includegraphics[height = 2.25 in, width = 3 in]{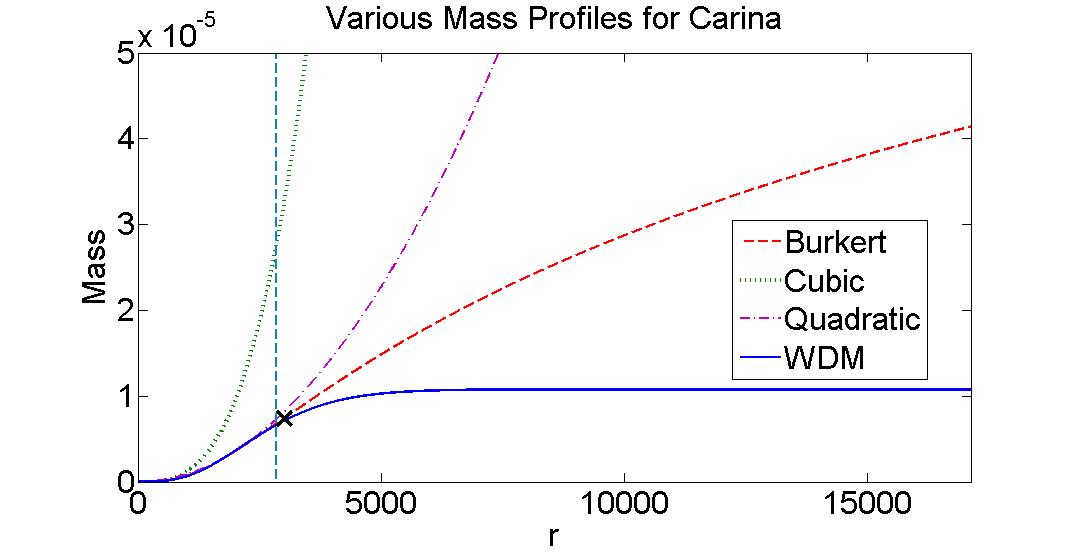}
    \includegraphics[height = 2.25 in, width = 3 in]{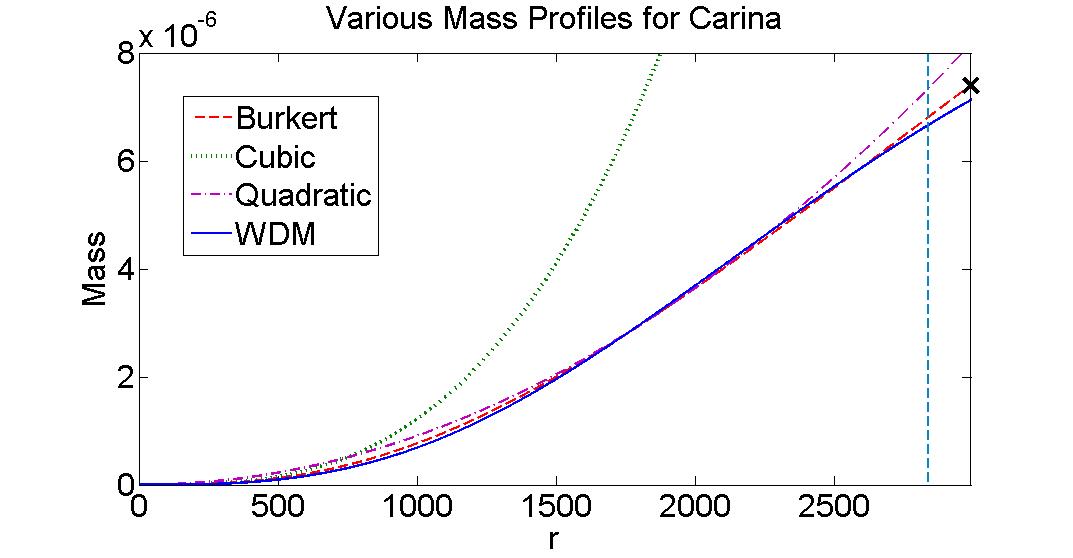}
  \end{center}

  \caption{Left: Plot of the Burkert mass profile for the Carina galaxy found by Salucci et al.\ \cite{Salucci11} along with a mass plot of a wave dark matter static ground state, the cubic function which is the leading term of the Taylor expansion of the Burkert mass profile, and the quadratic power function $\dfrac{M_{B}(r_{c})}{r_{c}^2}r^{2}$ where $r_{c}$ is the core radius of the Carina galaxy.  The $\mathbf{\times}$ marks the location of the inflection point of the Burkert mass profile, while the vertical line denotes the location of the outermost data point for the Carina galaxy and is presented for reference purposes only. Right: Closeup of the plot on the left over the $r$ interval $[0,r_{ip}]$.}

  \label{BurEKG_cmpr}

\end{figure}

In fact, several other models for dark matter mass profiles have similar initial behavior to the Burkert profile including a quadratic mass profile (which is not physical and is only included for the sake of comparisons) and wave dark matter mass profiles.  In Figure~\ref{BurEKG_cmpr}, we have collected several mass models that have similar behavior inside $r = r_{ip}$ to the Burkert mass profile computed by Salucci et al.\ for the Carina galaxy \cite{Salucci11}.  While these models have similar behavior inside $r = r_{ip}$, they are very different outside $r = r_{ip}$.


\begin{figure}

  \begin{center}
    \includegraphics[height = 1.75 in, width = 2 in]{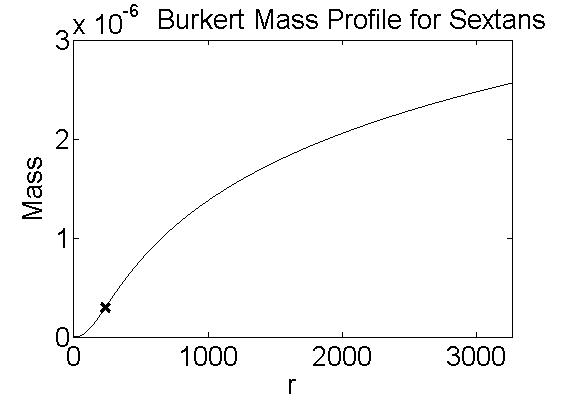}
    \includegraphics[height = 1.75 in, width = 2 in]{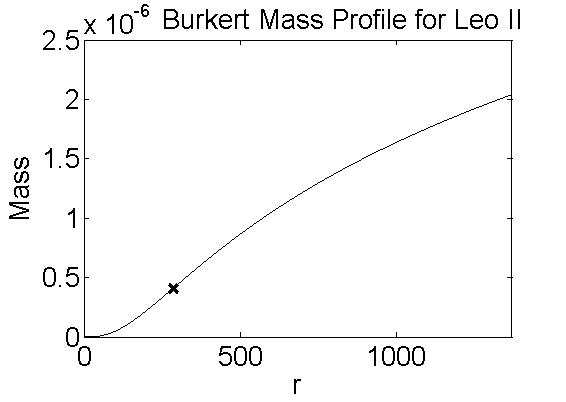}
    \includegraphics[height = 1.75 in, width = 2 in]{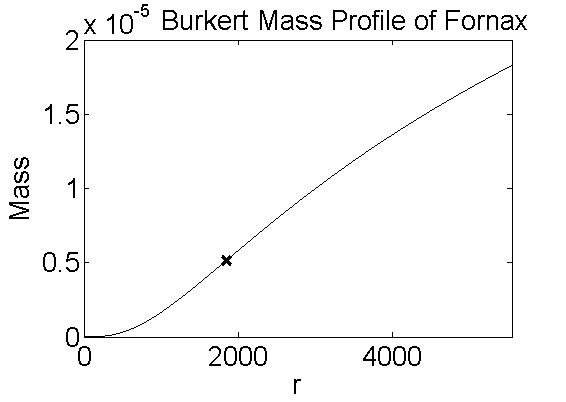}

    \includegraphics[height = 1.75 in, width = 2 in]{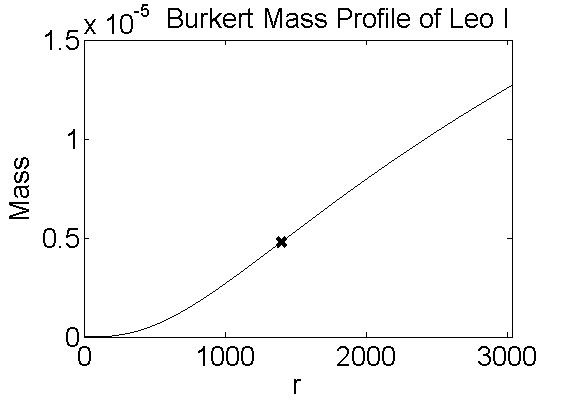}
    \includegraphics[height = 1.75 in, width = 2 in]{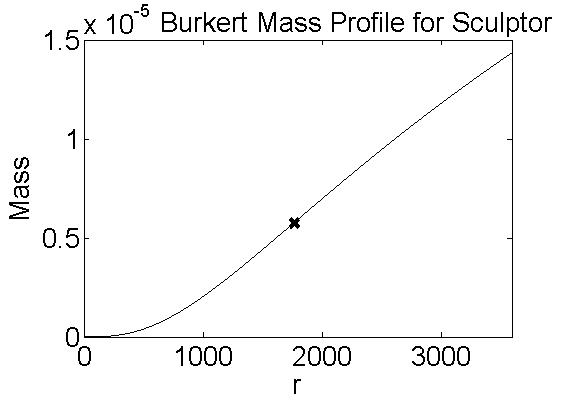}
    \includegraphics[height = 1.75 in, width = 2 in]{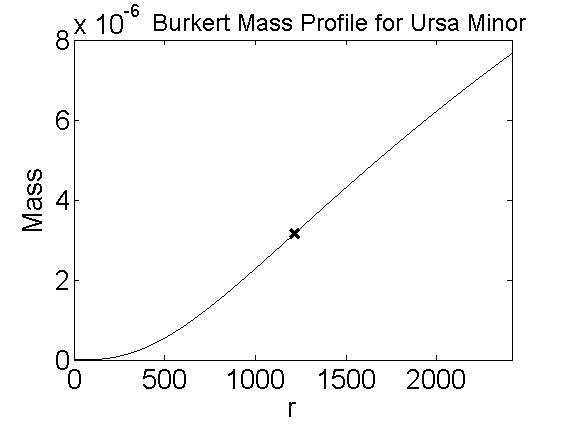}

    \includegraphics[height = 1.75 in, width = 2 in]{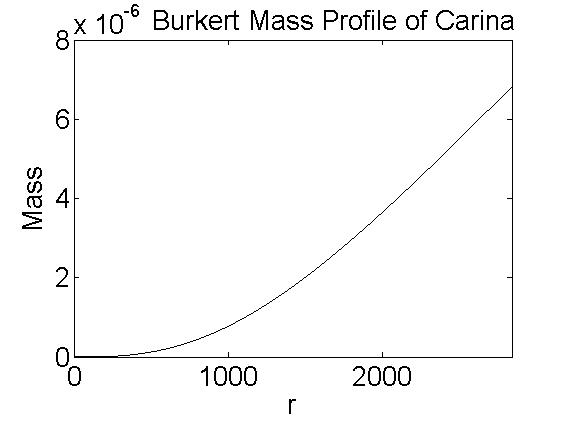}
    \includegraphics[height = 1.75 in, width = 2 in]{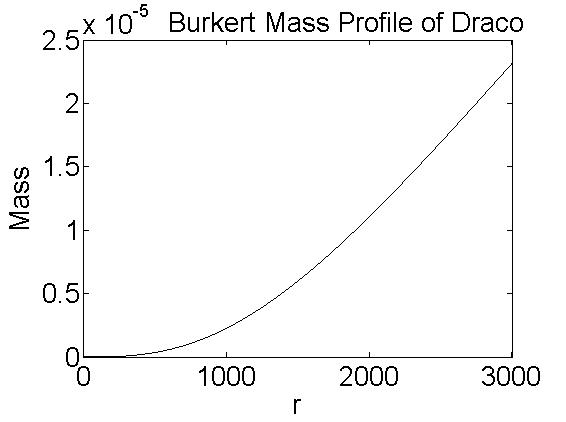}
  \end{center}

  \caption{Plots of the Burkert mass profiles computed by Salucci et al.\ of the eight classical dwarf spheroidal galaxies within the range of observable data.  The inflection point is marked on each plot by an $\mathbf{\times}$.  Carina and Draco have no inflection point marked because the inflection point for their Burkert mass profiles occurs outside the range of observable data.}

  \label{clsdSph_ip}

\end{figure}

We have computed the inflection points of each of the Burkert mass profiles computed by Salucci et al.\ for the eight classical dwarf spheroidal galaxies \cite{Salucci11} and have marked these points on a plot of each Burkert mass profile in Figure \ref{clsdSph_ip}.  We have constrained the viewing window of each plot to the range of data points collected.  That is, we plot the Burkert mass profiles on the interval $[0,r_{last}]$, where $r_{last}$ is the radius of the outermost data point given by Walker et al.\ \cites{Walker09,Walker10} for the observed velocity dispersion profiles.  We have presented them in order from greatest to least according to the ratio of $r_{last}/r_{ip}$.

In Table \ref{Bur-tab}, we have collected the defining parameters, $\rho_{0}$ and $r_{c}$, computed by Salucci et al.\ for the Burkert mass profiles which best predict the velocity dispersion profiles of each galaxy \cite{Salucci11}.  We have also collected the outermost data point, $r_{last}$, of these velocity dispersion profiles \cites{Walker09,Walker10}, as well as our computations of the inflection point, $r_{ip}$, and the ratio $r_{last}/r_{ip}$ for each of the classical dwarf spheroidal galaxies.  All quantities have been converted to geometrized units (the universal gravitational constant and the speed of light set to one) of (light)years for mass, length, and time.

\begin{table}

  \begin{center}

    \begin{tabular}{c|ccccc}
      Galaxy Name & $\rho_{0}$ ($\text{yr}^{-2}$) & $r_{c}$ (yr)         & $r_{last}$ (yr) & $r_{ip}$ (yr) & $r_{last}/r_{ip}$ \\
      \hline & & & \\
      Sextans     & $2.47 \times 10^{-14}$        & $1.53 \times 10^{2}$ & $3.26 \times 10^{3}$   & $2.32 \times 10^{2}$ & $14.05$ \\
      Leo II      & $1.83 \times 10^{-14}$        & $1.88 \times 10^{2}$ & $1.37 \times 10^{3}$   & $2.86 \times 10^{2}$ & $4.80$ \\
      Fornax      & $8.57 \times 10^{-16}$        & $1.21 \times 10^{3}$ & $5.54 \times 10^{3}$   & $1.84 \times 10^{3}$ & $3.01$ \\
      Leo I       & $1.83 \times 10^{-15}$        & $9.19 \times 10^{2}$ & $3.03 \times 10^{3}$   & $1.40 \times 10^{3}$ & $2.17$ \\
      Sculptor    & $1.10 \times 10^{-15}$        & $1.16 \times 10^{3}$ & $3.59 \times 10^{3}$   & $1.76 \times 10^{3}$ & $2.04$ \\
      Ursa Minor  & $1.83 \times 10^{-15}$        & $8.01 \times 10^{2}$ & $2.41 \times 10^{3}$   & $1.22 \times 10^{3}$ & $1.98$ \\
      Carina      & $2.90 \times 10^{-16}$        & $1.97 \times 10^{3}$ & $2.84 \times 10^{3}$   & $2.99 \times 10^{3}$ & $0.95$ \\
      Draco       & $8.19 \times 10^{-16}$        & $2.11 \times 10^{3}$ & $3.00 \times 10^{3}$   & $3.20 \times 10^{3}$ & $0.94$
    \end{tabular}

  \end{center}

  \caption{Burkert mass profile data for the eight classical dwarf spheroidal galaxies converted to units of years for mass, length, and time.  The parameters $\rho_{0}$ and $r_{c}$ are those found by Salucci et al.\ for the best fit Burkert profiles \cite{Salucci11}, and $r_{last}$ is the radius of the outermost data point given by Walker et al.\ \cites{Walker09,Walker10}.  Also included is the value of the inflection point, $r_{ip}$, of the Burkert mass profile for each galaxy and the ratio of $r_{last}$ to $r_{ip}$.}

  \label{Bur-tab}

\end{table}

\section{Static States of Wave Dark Matter}

Now that we have presented mass profiles which model actual data well, we need to describe the wave dark matter models we will use to make our comparison.  In the following, we present only the basic background information required to understand the model we use and refer the reader to  \cites{Bray10,Bray12} for more discussion on its motivation and successes thus far.

Let $(N,g)$ be a spacetime whose metric has signature $(-+++)$.  Let $f:N \to \C$ be a smooth complex-valued scalar field defined on the spacetime.  Finally, let $f$ and $g$ satisfy the Einstein-Klein-Gordon equations,
\begin{subequations}\label{EKG}
    \begin{align}
      \label{EKG-1} G &= 8\pi \mu_{0}\pnth{\frac{df \otimes d\cf + d\cf \otimes df}{\Up^{2}} - \pnth{\frac{\abs{df}^{2}}{\Up^{2}} + \abs{f}^{2}}g} \\
      \label{EKG-2} \glap{g}f &= \Up^2 f
    \end{align}
\end{subequations}
where $\glap{g}$ is the Laplacian with respect to the metric $g$.  The parameter $\Up$ is a fundamental constant of this system and its value must be determined in order to use these equations to model dark matter in the universe.  On the other hand, the parameter $\mu_{0}$ is not fundamental to the system and can be completely absorbed into $f$ if desired.

In  \cite{Bray10}, wave dark matter may be modeled as a real scalar field or as a complex scalar field.  For the Einstein-Klein-Gordon system, a complex scalar field is equivalent to two real scalar fields.  Complex scalar fields are more convenient because they have static spacetime solutions.  Analogous ``nearly static'' real solutions can be achieved by redefining $f$ for a static spacetime solution as $\sqrt{2} \, \Re(f)$ whose spacetime metrics are very similar except for a high frequency oscillating pressure term which averages out to zero.  Preliminary estimates suggest that this oscillating pressure effect might not be significant enough to make it measurable on physically relevant time scales.  Hence, there may not be a testable difference between the predictions of real and complex scalar field dark matter at this time, though more thought on this question is well deserved.  In either case, we will call these scalar field models ``wave dark matter.''

Since we wish to compare solutions of (\ref{EKG}) to the spherically symmetric Burkert mass profiles computed by Salucci et al.\ \cite{Salucci11}, we choose to work in spherical symmetry.  In a recent paper \cite{Parry12-1}, Parry surveyed the well-known form of the metric of a general spherically symmetric spacetime in polar-areal coordinates, namely,
\begin{equation} \label{metric}
  g = -\e^{2V(t,r)}\, dt^{2} + \pnth{1 - \frac{2M(t,r)}{r}}^{-1}\, dr^{2} + r^{2}\, d\si^{2},
\end{equation}
for real valued functions $V$ and $M$ and where $d\si^{2} = d\theta^{2} + \sin^{2}\theta\, d\varphi^{2}$ is the standard metric on the unit sphere.  This metric has the following useful properties.  The function $M(t,r)$ is the Hawking mass of the metric sphere of radius $r$ and time $t$.   Under the Einstein equation, $G = 8\pi T$, $M$ is also the flat volume integral of the energy density term in the stress energy tensor.  This motivates interpreting the function $M(t,r)$ as the mass inside the metric sphere of radius $r$ at time $t$.  Finally, given the Einstein equation, in the low field limit, $V$ is approximately the gravitational potential of the system.  We refer the reader to  \cite{Parry12-1} for detailed proofs of these facts.

It is also shown in  \cite{Parry12-1} that in spherical symmetry and using the metric (\ref{metric}), solving the Einstein-Klein-Gordon system (\ref{EKG}) reduces to solving the system
\begin{subequations} \label{NCpde}
  \begin{align}
    \label{NCpde1a} M_{r} &= 4\pi r^{2}\mu_{0}\pnth{\abs{f}^{2} + \pnth{1 - \frac{2M}{r}}\frac{\abs{f_{r}}^{2} + \abs{p}^{2}}{\Up^{2}}} \\
    \label{NCpde2a} V_{r} &= \pnth{1 - \frac{2M}{r}}^{-1}\pnth{\frac{M}{r^{2}} - 4\pi r\mu_{0}\pnth{\abs{f}^{2} - \pnth{1 - \frac{2M}{r}}\frac{\abs{f_{r}}^{2} + \abs{p}^{2}}{\Up^{2}}}} \displaybreak[0]\\
    \label{NCpde3a} f_{t} &= p\e^{V}\sqrt{1 - \frac{2M}{r}} \\
    \label{NCpde4a} p_{t} &= \e^{V}\pnth{-\Up^{2}f\pnth{1 - \frac{2M}{r}}^{-1/2} + \frac{2f_{r}}{r}\sqrt{1 - \frac{2M}{r}}} + \p_{r}\pnth{\e^{V}f_{r}\sqrt{1 - \frac{2M}{r}}}.
  \end{align}
\end{subequations}

To solve this system, we need boundary conditions.  At the central value, we require all of the functions to be smooth.  Since all of the functions are spherically symmetric, this implies that $M_{r}$, $V_{r}$, $f_{r}$, and $p_{r}$ all vanish at $r=0$ for all $t$.  We will also require that the spacetime be asymptotically Schwarzschild, that is, it approaches a Schwarzschild metric as $r \to \infty$.  Specifically, this implies that
\begin{align}
  \label{AB2a} \e^{2V} &\to \ka^{2}\pnth{1 - \frac{2M}{r}} \quad \text{as } r \to \infty \\
  \label{AB1a} \glap{g_{S}}f &\to \Upsilon^2 f \quad \text{and} \quad f \to 0 \quad \text{as } r \to \infty
\end{align}
where $\ka > 0$ and $g_{S}$ is the appropriate Schwarzschild metric.  Since $f \to 0$ as $r \to \infty$, $M$ approaches a constant value $m$, which is the total mass of the system.

Note that these boundary conditions ensure that as $r \to \infty$, the metric $g$ in equation (\ref{metric}) becomes the Schwarzschild metric
\begin{equation}\label{SW-metric}
  g_{S} = -\ka^{2}\pnth{1 - \frac{2m}{r}}\, dt^{2} + \pnth{1 - \frac{2m}{r}}^{-1}\, dr^{2} + r^{2}\, d\si^{2}.
\end{equation}
Thus $\ka$ represents a scaling of the $t$ coordinate in the standard Schwarzschild metric.  The effect of this on our discussion is that $V \to \ln \ka$ as $r \to \infty$.  We have to numerically solve these equations and so in practice, we will impose these boundary conditions at an artificial right hand boundary point $r_{max}$ and solve the system on the $r$-interval $[0,r_{max}]$.

One of the simplest solutions to this system are those where the scalar field is of the form
\begin{equation}\label{sfstaticstate}
  f(t,r) = \e^{i\om t}F(r)
\end{equation}
where $\om \in \RR$ is a constant and $F$ is real valued.

Note that for $f$ of this form, solving equation (\ref{AB1a}) for large $r$ and requiring the solution to decay to $0$ yields that, for large $r$, $F$ must satisfy
\begin{equation}
  F' + \pnth{\sqrt{\Up^{2} - \frac{\om^{2}}{\ka^{2}}} + \frac{1}{r}}F \approx 0.
\end{equation}
Requiring this condition on our system ensures that $f$ appropriately decays to $0$ as $r \to \infty$ \cite{Parry12-3}.

Solutions of the form in equation (\ref{sfstaticstate}) produce static metrics and, once substituted into the system (\ref{NCpde}), yield the following set of ODEs \cite{Parry12-3},
\begin{subequations} \label{NCpdec}
  \begin{align}
    \label{NCpde1c} M' &= 4\pi r^{2}\mu_{0}\brkt{\pnth{1 + \frac{\om^{2}}{\Up^{2}}\e^{-2V}}\abs{F}^{2} + \pnth{1 - \frac{2M}{r}}\frac{\abs{H}^{2}}{\Up^{2}}} \\
    \label{NCpde2c} V' &= \pnth{1 - \frac{2M}{r}}^{-1}\brce{\frac{M}{r^{2}} - 4\pi r\mu_{0}\brkt{\pnth{1 - \frac{\om^{2}}{\Up^{2}}\e^{-2V}}\abs{F}^{2} - \pnth{1 - \frac{2M}{r}}\frac{\abs{H}^{2}}{\Up^{2}}}} \\
    \label{NCpde3c} F' &= H \\
    \label{NCpde4c} H' &= \pnth{1 - \frac{2M}{r}}^{-1}\brkt{\pnth{\Up^{2} - \frac{\om^{2}}{\e^{2V}}} F + 2H\pnth{\frac{M}{r^{2}} + 4\pi r\mu_{0}\abs{F}^{2} - \frac{1}{r}}}
  \end{align}
\end{subequations}
with boundary conditions
\begin{align}
  \label{cenval} F(0) &= 1, & H(0) &= 0, & M(0) &= 0, & V(0) &= V_{0},
\end{align}
\begin{align}
  \label{AB1c} F'(r_{max}) + \pnth{\sqrt{\Up^{2} - \frac{\om^{2}}{\ka^{2}}} + \frac{1}{r_{max}}}F(r_{max}) &\approx 0, \\
  \label{AB2c} V(r_{max}) - \frac{1}{2}\ln\pnth{1 - \frac{2M(r_{max})}{r_{max}}} - \ln \ka &\approx 0,
\end{align}
by equations (\ref{AB2a}) and (\ref{AB1a}).  For simplicity, we set $\ka = 1$, which corresponds to the assumption on our choice of $t$ coordinate that $V$ goes to zero at infinity.  A solution to these equations depends on the choice of the parameters $\Up$, $\mu_{0}$, $\om$, and $V_{0}$.  We solve a shooting problem for $\om$ and $V_{0}$ to satisfy (\ref{AB1c}) and (\ref{AB2c}) leaving $\Up$ and $\mu_{0}$ freely selectable.

\begin{figure}

  \begin{center}
    \includegraphics[height = 2.25 in, width = 3 in]{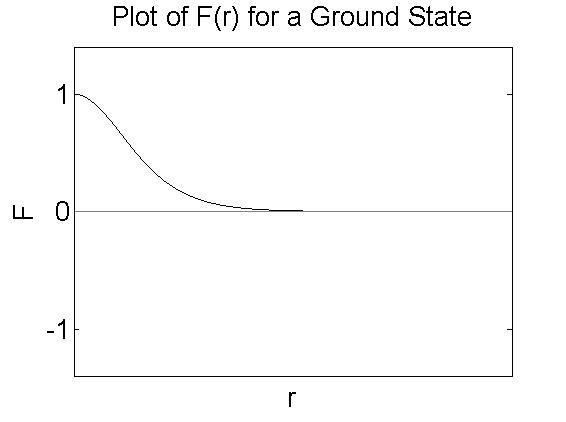}
    \includegraphics[height = 2.25 in, width = 3 in]{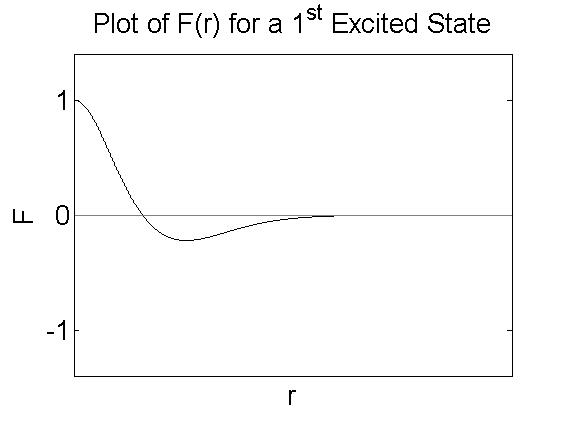}

    \includegraphics[height = 2.25 in, width = 3 in]{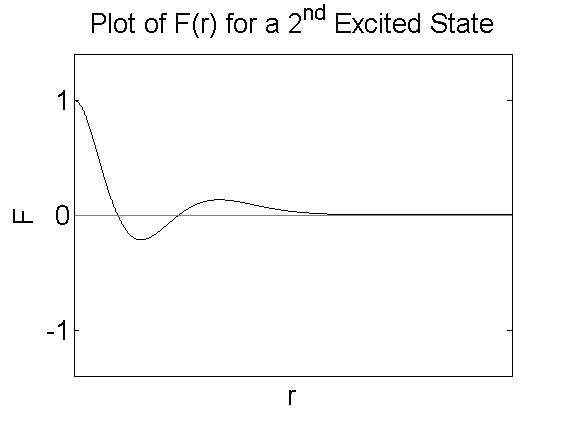}
    \includegraphics[height = 2.25 in, width = 3 in]{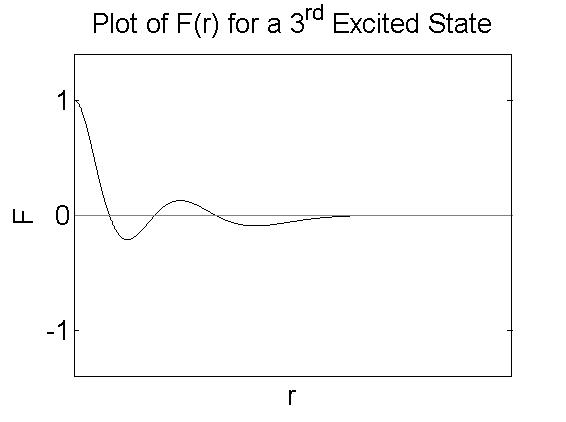}
  \end{center}

  \caption{Plots of spherically symmetric static state scalar fields (specifically the function $F(r)$ in (\ref{sfstaticstate})) in the ground state and first, second, and third excited states.  Note the number of nodes (zeros) of each function.}

  \label{state_ex}

\end{figure}

\begin{figure}

    \begin{center}
        \includegraphics[height = 2.25 in, width = 3 in]{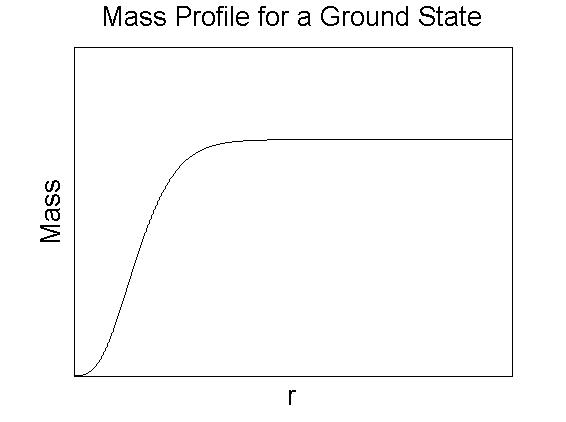}
        \includegraphics[height = 2.25 in, width = 3 in]{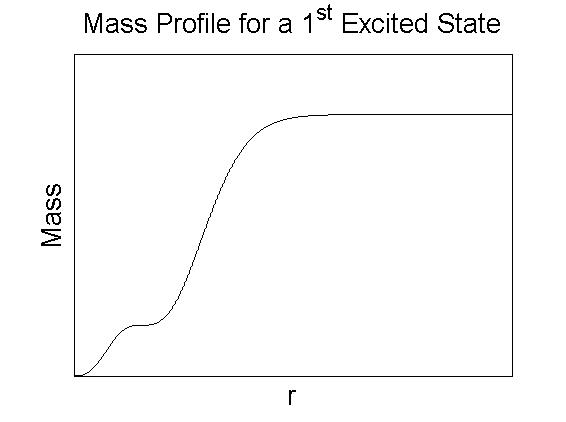}

        \includegraphics[height = 2.25 in, width = 3 in]{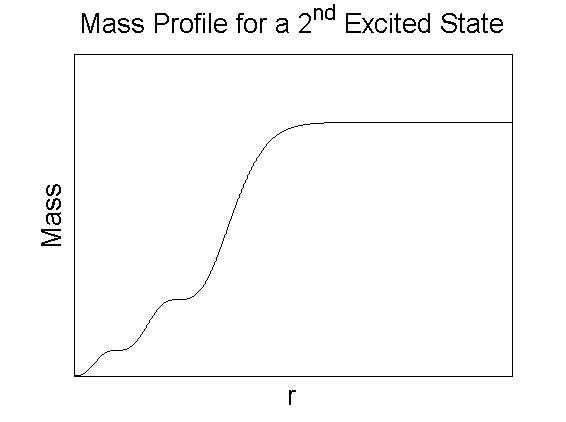}
        \includegraphics[height = 2.25 in, width = 3 in]{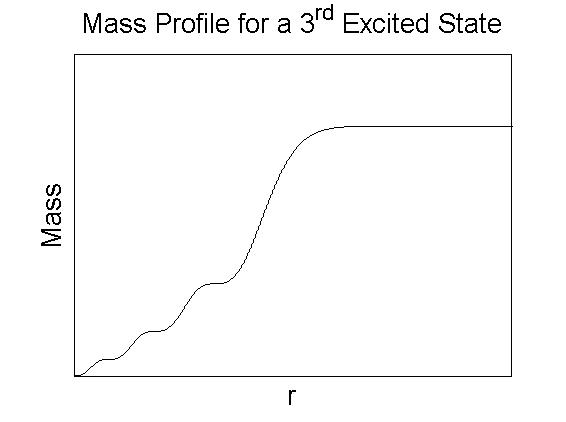}
    \end{center}

    \caption{Mass profiles for a static ground state and first, second, and third excited states of wave dark matter.}

    \label{state_m}

\end{figure}

For each choice of $\Up$ and $\mu_{0}$, there are an infinite number of discrete finite mass solutions characterized by the number of zeros that $F$ exhibits \cite{Parry12-3}.  These are called static states.  A static state with no zeros is called a ground state.  With $n$ zeros for $n>0$, it is called an $n^{\text{th}}$ excited state.  In Figure \ref{state_ex}, we have plotted examples of $F$ for a ground through third excited state.  In Figure \ref{state_m}, we have presented the plots of the mass, $M$, corresponding to the plots of $F$ in Figure \ref{state_ex}.

\subsection{Working Value of \texorpdfstring{$\Up$}{Upsilon}}

\begin{figure}

  \begin{center}
    \includegraphics[height = 1.75 in, width = 2 in]{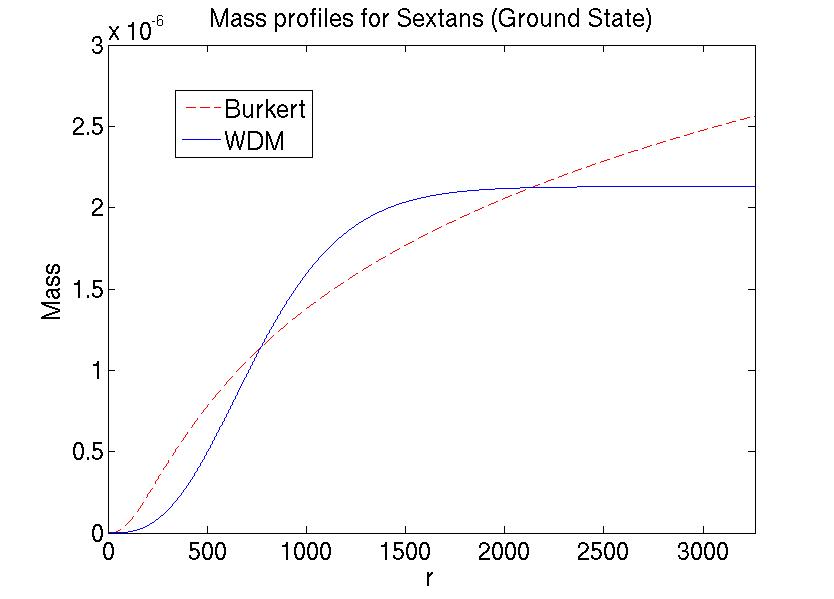}
    \includegraphics[height = 1.75 in, width = 2 in]{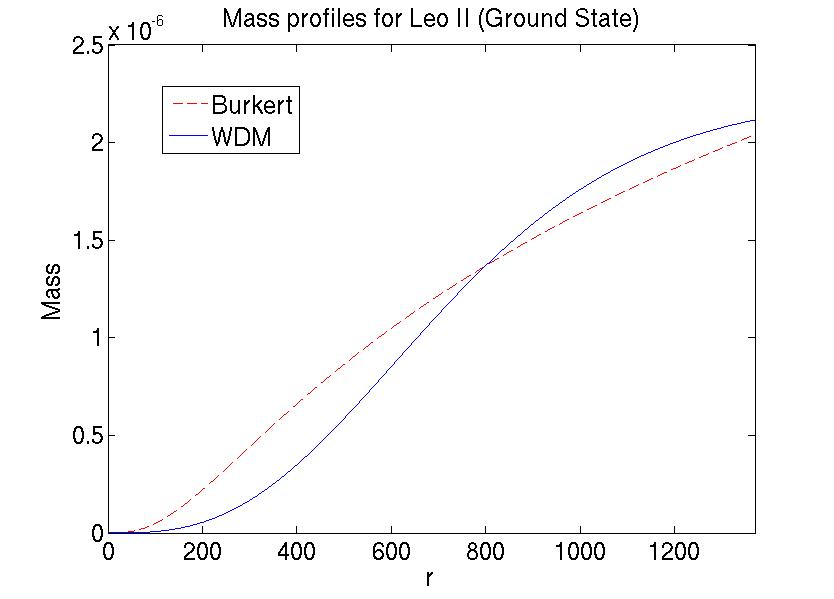}
    \includegraphics[height = 1.75 in, width = 2 in]{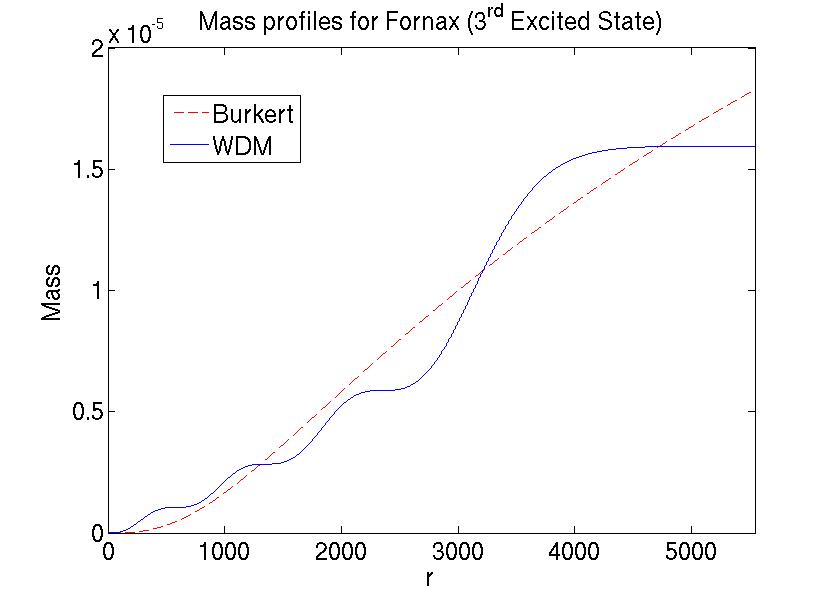}

    \includegraphics[height = 1.75 in, width = 2 in]{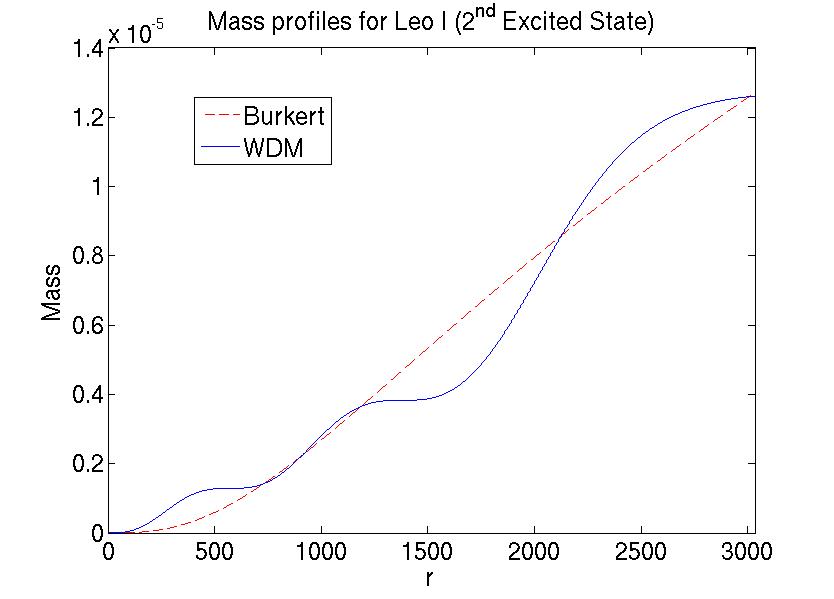}
    \includegraphics[height = 1.75 in, width = 2 in]{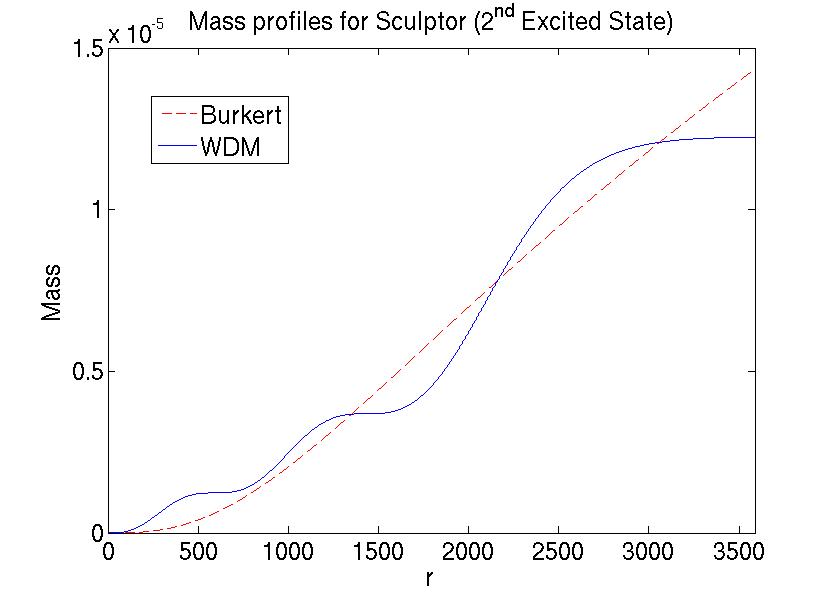}
    \includegraphics[height = 1.75 in, width = 2 in]{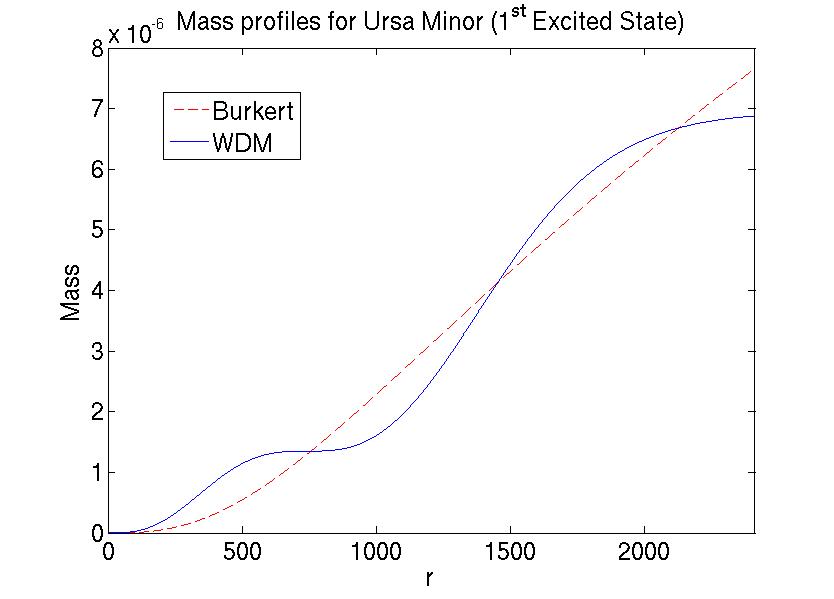}

    \includegraphics[height = 1.75 in, width = 2 in]{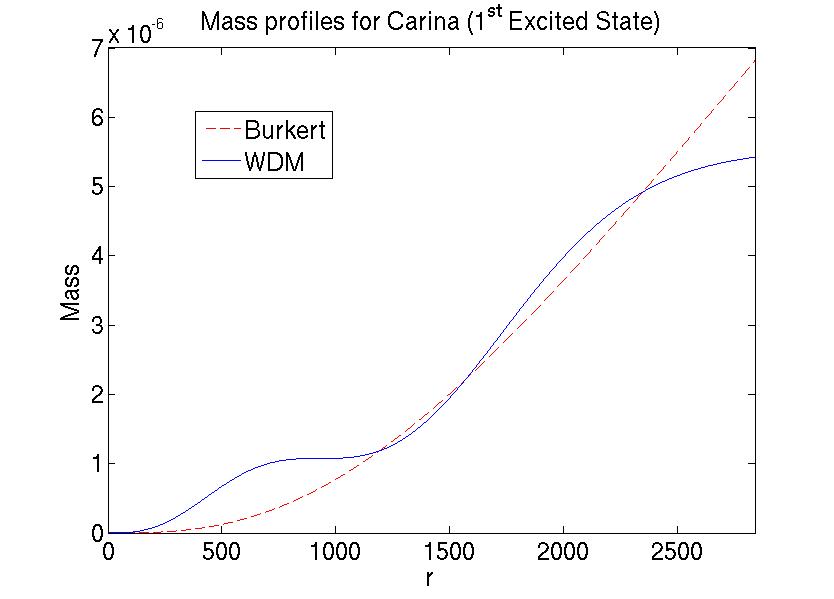}
    \includegraphics[height = 1.75 in, width = 2 in]{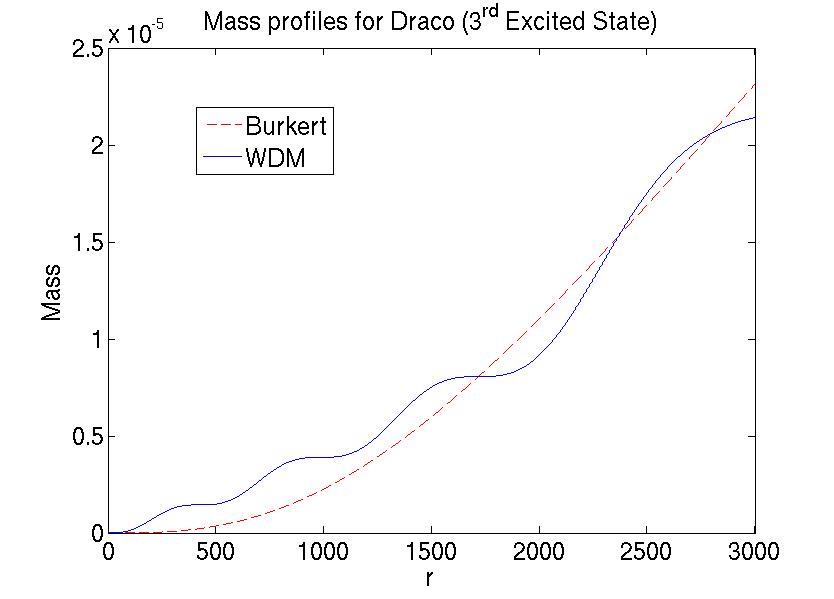}
  \end{center}

  \caption{Static state mass profiles for $\Up = 50$ which are each a best fit to the Burkert profiles of the corresponding  dwarf spheroidal galaxy.  For $\Up = 50$, we picked an $n^{\text{th}}$ excited state whose best fit profile matched the Burkert profile qualitatively well.  This shows that $\Up = 50$ is a reasonable working value of $\Up$.  However, it does not imply that the actual value of $\Up$ is $50$ or that these galaxies are correctly modeled by the presented $n^{\text{th}}$ excited state.  The units on $\Up$ are $\text{yr}^{-1}$.}

  \label{Ups50}

\end{figure}

As stated before, our goal is to find a value of $\Up$ that is compatible with the Burkert mass profiles that Salucci et al.\ \cite{Salucci11} computed to model the dark matter in the eight classical dwarf spheroidal galaxies.  For $\Up = 50 \text{ yr}^{-1}$, there is at least one wave dark matter static state that matches the Burkert mass profiles reasonably well.  We have plotted such matches in Figure \ref{Ups50}.  Thus we have chosen to use
\begin{equation}\label{Upwrkval}
  \Up = 50 \text{ yr}^{-1}
\end{equation}
as a working value of $\Up$ in our future work with wave dark matter until we have the capability to make a more accurate approximation or precise measurement of this value.

While we have chosen $\Up = 50 \text{ yr}^{-1}$ as a working value of $\Up$ since it corresponds to wave dark matter models compatible with other well fitting models, the above does not constitute a precise measurement of the value of $\Up$.  In the remainder of this paper, we will show that, under precise assumptions, the value of $\Up$ can be bounded above, a first step towards obtaining a direct measurement of $\Up$.  To do this, we first describe a few extra properties of these spherically symmetric static states.

\subsection{Hyperbolas of Constant \texorpdfstring{$\Up$}{Upsilon}}

Parry showed in \cite{Parry12-3} how the parameters which define an $n^{\text{th}}$ excited state, $\Up$, $\mu_{0}$, $\om$, and $V_{0}$, as well as the values of the total mass, $m$, and the radius, $r_{h}$, called the half mass radius, for which $M(r_{h}) = m/2$, are related to each other.  The approximations which follow only apply to solutions in the long wavelength, low field limit, which is what is relevant for modeling galaxies.  In particular, for constant $\Up$, there is a one parameter family of solutions for each $n^{\text{th}}$ excited state defined by the equations
\begin{align}
  \label{Scorr1} \om^{n}(\Up,\mu_{0}) &\approx \Up \exp\pnth{C_{frequency}^{n}\frac{\sqrt{\mu_{0}}}{\Up}} \\
  \label{Scorr2} V_{0}^{n}(\Up,\mu_{0}) &\approx C_{potential}^{n}\frac{\sqrt{\mu_{0}}}{\Up} \\
  \label{Scorr3} m^{n}(\Up,\mu_{0}) &\approx C_{mass}^{n}\Up^{-3/2}\mu_{0}^{1/4} \\
  \label{Scorr4} r_{h}^{n}(\Up,\mu_{0}) &\approx C_{radius}^{n}\Up^{-1/2}\mu_{0}^{-1/4}.
\end{align}
where the constants $C_{\ast}^{n}$ depend upon which state we wish to consider (i.e. they depend on $n$).  Thus for constant $\Up$, the different possible excited states satisfying (\ref{cenval}), (\ref{AB1c}), and (\ref{AB2c}) are defined entirely by the value of $\mu_{0}$, which corresponds to the value of the energy density function, $\mu$, as defined in \cite{Parry12-1} at the origin via the equation
\begin{equation}
  \mu(0) = \mu_{0}\pnth{1 + \frac{\om^{2}}{\Up^{2}}\e^{-2V_{0}}}.
\end{equation}

The equations relevant to our discussion here are (\ref{Scorr3}) and (\ref{Scorr4}).  We have collected the values of $C_{mass}^{n}$ and $C_{radius}^{n}$ from these two equations for the ground through fifth excited states as well as for the tenth and twentieth excited states in Table \ref{ScorrVal}.  A complete table of the values of all the constants $C_{\ast}^{n}$ for these excited states can be found in \cite{Parry12-3}.

\begin{table}

  \begin{center}

  \begin{tabular}{c|cc}
    $n$ & $C_{mass}^{n}$ & $C_{radius}^{n}$ \\
    \hline & & \\
    $0$ & $4.567 \pm 0.05$ & $0.8462 \pm 0.004$ \\
    $1$ & $10.22 \pm 0.10$ & $2.2894 \pm 0.009$ \\
    $2$ & $15.81 \pm 0.16$ & $3.8253 \pm 0.014$ \\
    $3$ & $21.37 \pm 0.22$ & $5.3994 \pm 0.018$ \\
    $4$ & $26.91 \pm 0.27$ & $6.9860 \pm 0.022$ \\
    $5$ & $32.42 \pm 0.33$ & $8.5606 \pm 0.026$ \\
    $10$ & $60.32 \pm 1.18$ & $15.1357 \pm 0.039$ \\
    $20$ & $116.62 \pm 2.57$ & $29.6822 \pm 0.107$
  \end{tabular}

  \caption{Values of the constants in equations (\ref{Scorr3}) and (\ref{Scorr4}) for the ground through fifth excited states as well as the tenth and twentieth excited states.  We have given these values error ranges which encompass the interval we observed in our experiments, but it is possible that values outside our ranges here could be observed.  However, we do not expect them to be outside by much if the discretization of $r$ used in solving the ODEs is sufficiently fine.  Note also that our values have less precision as we increase $n$.  This is because as $n$ increases, it becomes more difficult to compute the states with as much precision.}

  \label{ScorrVal}

  \end{center}

\end{table}

\begin{figure}

    \begin{center}
      \includegraphics[width=3 in]{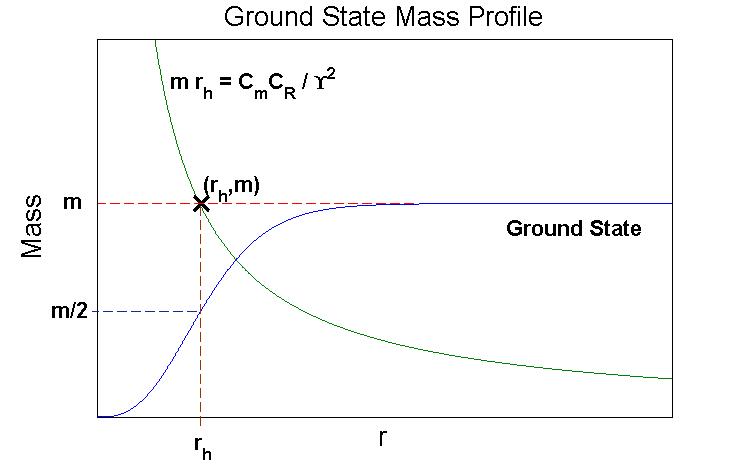}
      \includegraphics[width=3 in]{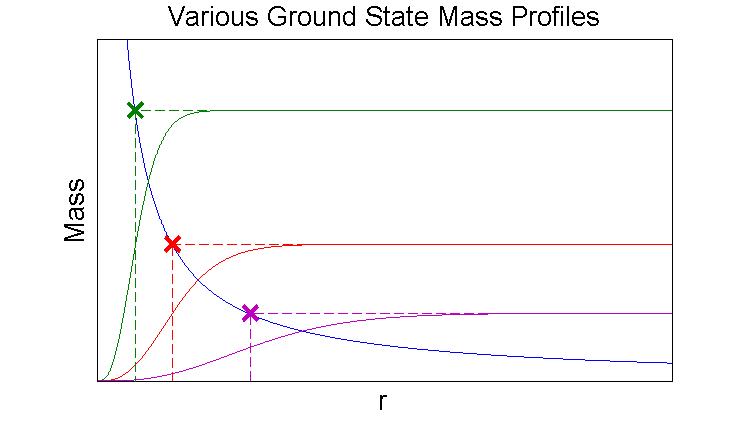}
    \end{center}

    \caption{Left: Plot of the mass profile of a ground state with its corresponding hyperbola of constant $\Up$ overlayed.  Any ground state mass profile that keeps the presented relationship with this hyperbola corresponds to the same value of $\Up$.  Right: Examples of different ground state mass profiles corresponding to the same value of $\Up$.  The corresponding hyperbola of constant $\Up$ is overlayed.  Notice that all three mass profiles have the same relationship with the hyperbola.}

    \label{mass-hyperbola-figure}

\end{figure}

As explained in  \cite{Parry12-3}, equations (\ref{Scorr3}) and (\ref{Scorr4}) imply that the product of $m$ and $r_{h}$ does not depend on the value of $\mu_{0}$, but only on $\Up$.  Specifically,
\begin{equation}\label{mass-hyperbola}
    m r_{h} = \frac{C_{mass}C_{radius}}{\Up^{2}},
\end{equation}
where we have suppressed the notation of $n$.  If $\Up$ is constant, then, because both $C_{mass}$ and $C_{radius}$ are positive, the right hand side of this equation is some positive constant, $k$, and we have
\begin{equation}
  m r_{h} = k
\end{equation}
which defines a hyperbola.  Thus, for a given $n^{\text{th}}$ excited state, all of the possible mass profiles for a constant value of $\Up$ lie along a hyperbola.  We illustrate this phenomenon in Figure \ref{mass-hyperbola-figure}.

\subsection{Fitting Burkert Mass Profiles}

With these properties of static state mass profiles in mind, we turn our attention to finding static state mass profiles that best fit the Burkert mass profiles computed by Salucci et al.\ \cite{Salucci11}  Given a Burkert mass profile, $M_{B}$, a value for $\Up$, and a specific state (i.e. value for $n$), we define for our purposes the best fit wave dark matter static state mass profile, $M_{W}$, as the one which minimizes the $L^{2}$ norm of the difference between these profiles, $E$, given by
\begin{equation}\label{L2Norm-exact}
    E = \norm{M_{B} - M_{W}}_{L^{2}}^{2} = \int_{0}^{r_{last}} (M_{B} - M_{W})^{2}\, dr.
\end{equation}
Of course, since we compute the static states numerically, we have to approximate this norm by an appropriate Riemann sum defined on a discretization of the interval $[0,r_{last}]$.

To find this minimum, we first note that since $\Up$ is fixed, we can write the total mass $m$ and the value of $\mu_{0}$ in terms of a choice of $r_{h}$ via equations (\ref{mass-hyperbola}) and (\ref{Scorr4}) respectively.  Thus, we parameterize the different mass profiles of constant $\Up$, and hence $E$ by $r_{h}$, that is, $E = E(r_{h})$.  Furthermore, since all of the static state mass profiles of constant $\Up$ lie on a hyperbola, there will be a
value of $r_{h}$ that yields the minimum of $E(r_{h})$.

To make computing the best fits more uniform from galaxy to galaxy, we make the choice $r_{h} = b r_{c}$, where $b>0$ and $r_{c}$ is the core radius of the Burkert profile we wish to match, and vary the free parameter $b$.  To compute which value of $b$ produces a minimum of value of $E(r_{h})$, we create a grid of $r_{h}$ values around an initial choice of $b$ of the form $[(b-step)r_{c}, br_{c}, (b+step)r_{c}]$ for some $step>0$.  Next we compute $E(r_{h})$ for each of the values of $r_{h}$ and shift the grid, if necessary, so that it is centered on the $r_{h}$ value which yielded the smallest value of $E(r_{h})$.  If the grid shifts, we recompute $E(r_{h})$ on the new grid and continue to shift, if necessary.  Once the minimum $E(r_{h})$ value occurs at the center of the grid, we keep that point as the center, but cut the step size in half.  We then run this shifting procedure again for this smaller grid until the minimum is at the center and then we shrink again.  We continue to shrink the step size until we get to a predetermined terminal value.  We generally would run the procedure until the step size was less than or equal to $2^{-10}$.

We have already seen generic examples of the results of this matching procedure as the plots in Figure \ref{Ups50} are best fits for their respective galaxies, given an $n^{\text{th}}$ excited state, and the value $\Up = 50 \text{ yr}^{-1}$.  However, this best fitting procedure also provides a method of finding values of $\Up$, for $\Up$ sufficiently large, which produce untenable matches to the Burkert profiles of Salucci et al.\ \cite{Salucci11}  This is the topic of the next section.

\subsection{Upper Bound for \texorpdfstring{$\Up$}{Upsilon}}

To find an upper bound for $\Up$, we first need to explain how the static states change as $\Up$ gets large.  Equation (\ref{mass-hyperbola}) implies that for a given $n^{\text{th}}$ excited state, as $\Up$ increases, the product $mr_{h}$ decreases.  The hyperbolas corresponding to smaller values of $m r_{h}$ are those that lie closer to the mass and radius axes.

\begin{figure}

  \begin{center}
    \includegraphics[width=3 in]{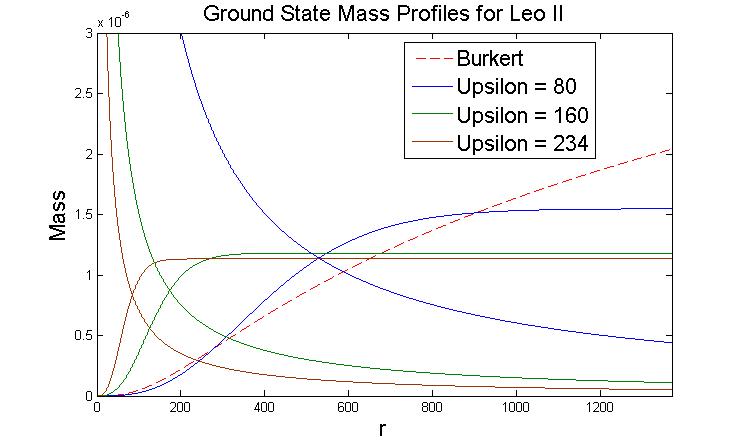}
    \includegraphics[width=3 in]{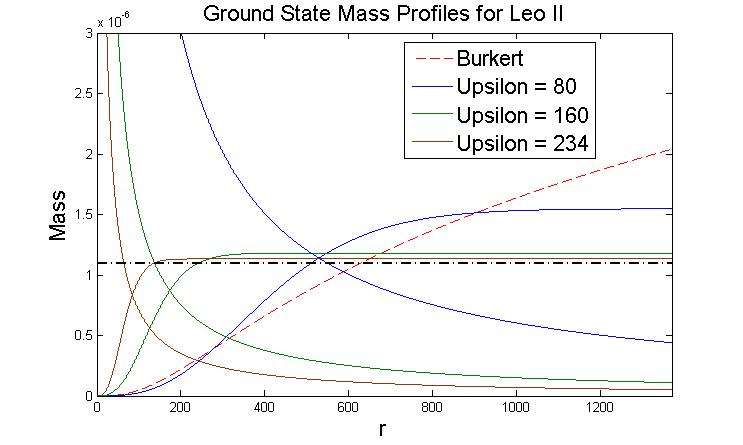}
  \end{center}

  \caption{Left: Ground state mass profiles of various values of $\Up$ that are best fits to the Burkert mass profile found by Salucci et al.\ \cite{Salucci11} for the Leo II galaxy.  The corresponding hyperbolas of constant $\Up$ on which these profiles lie are also plotted.  Ground states and their corresponding hyperbolas are drawn in the same color.  Right: The same plots as in the left frame, but with the constant function which best fits the Burkert profile also plotted.  Note that the best fit mass profiles approach this constant mass profile as $\Up$ increases.}

  \label{GroundStateMatchUpsUp}

\end{figure}

Now consider the $n^{\text{th}}$ excited state mass profile that is the best fit to a Burkert mass profile for a given $\Up$.  As $\Up$ increases, the hyperbola to which this static state mass profile corresponds will get closer to the mass and radius axes, but since the mass profile must also minimize $E(r_{h})$, the value of its mass will not tend to $0$.  Since $m r_{h}$ tends to zero as $\Up \to \infty$, it must be instead that $r_{h} \to 0$ as $\Up \to \infty$.  This implies that, as $\Up$ increases, more of the constant portion of the best fitting $n^{\text{th}}$ excited state mass profile will be compared to the Burkert profile.  Thus, as $\Up \to \infty$, the best fitting $n^{\text{th}}$ excited state mass profile will limit to the constant function of $r$ that best fits the Burkert profile under the same fitting criteria used for the static states.  We illustrate this phenomenon in Figure \ref{GroundStateMatchUpsUp}.

As $\Up \to \infty$, the initial increasing region (i.e. the region before the constant portion) of the $n^{\text{th}}$ excited state mass profile that best fits a Burkert mass profile becomes more compressed.  This initial region is where all of the dark matter mass is located.  Thus as $\Up$ increases, the dark matter corresponding to the best fit $n^{\text{th}}$ excited state extends out to smaller radii.  However, observations suggests that dwarf spheroidal galaxies are dark matter dominated at all observable radii \cite{Kleyna02}.  Thus the best fit $n^{\text{th}}$ excited state mass profiles for large $\Up$ do not represent observations well and can be rejected.  The question then is exactly when should we reject them.

Since every static state has the initial increasing region just described, the best fit $n^{\text{th}}$ excited state for any value of $\Up$ will be a better fit than the best fit constant function.  Moreover, since this initial region becomes more compressed as $\Up \to \infty$, for large $\Up$, the value of $E(r_{h})$ for the best fit $n^{\text{th}}$ excited state mass profile increases monotonically as $\Up \to \infty$ approaching the value of $E$ for the best fitting constant function.

This suggests a criteria for when to reject values of $\Up$.  We will reject a best fit $n^{\text{th}}$ excited state mass profile, and hence its corresponding value of $\Up$, as an untenable model of the dark matter mass if its value of $E(r_{h})$ is greater than some prescribed fraction of the value of $E$ for the best fitting constant function.  We chose to use $80\%$.  Explicitly, we use the following rejection criteria.

\begin{rej}\label{rejcrit_1}
  Given $\Up$, $n$, and a Burkert mass profile $M_{B}$, let $M_{W}$ be the spherically symmetric $n^{\text{th}}$ excited state mass profile corresponding to $\Up$ that best fits $M_{B}$, that is, that minimizes $E$ from equation (\ref{L2Norm-exact}) along the hyperbola defined by the value of $\Up$ and equation (\ref{mass-hyperbola}).  Let $E_{W}$ be the value of $E$ for this mass profile.  Furthermore, let $M_{C}$ be the constant function which best fits $M_{B}$, also by minimizing the corresponding function $E$, and let $E_{C}$ be the value of $E$ for the constant function.  Reject the given value $\Up$ as a tenable value for this fundamental constant if
    $$E_{W} \geq .8 E_{C}.$$
\end{rej}

In other words, any fit that is less than $20\%$ better than the best fitting constant function of $r$ is rejected as a bad fit.

\begin{figure}

  \begin{center}
    \includegraphics[height = 2.25 in]{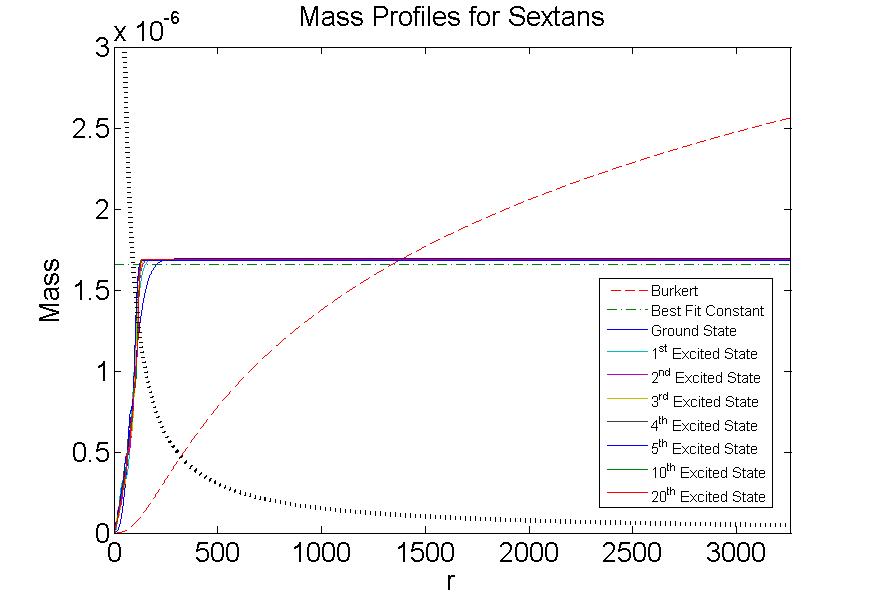}
  \end{center}

  \caption{The Burkert mass profile found by Salucci et al.\ \cite{Salucci11} for the Sextans galaxy.  The best fit static state mass profiles for a ground through fifth excited state, tenth excited state, and twentieth excited state all lying on the same hyperbola are overlayed on the plot.  The hyperbola here satisfies the rejection criteria for all of the different static states represented in the plot, thus all of these static states correspond to an upper bound on the value of $\Up$ for Sextans for their respective value of $n$ (i.e. the set of $n^{\text{th}}$ excited states).  Note how close together all of the states are.  This is due to the fact that the majority of their profiles which are being compared to the Burkert mass profile is the common and constant portion of the profiles.}

  \label{badfitSextans}

\end{figure}

For each of the eight classical dwarf spheroidal galaxies and $n \in \{0,1,2,3,4,5,10,20\}$, we computed values of $\Up$ that yielded $n^{\text{th}}$ excited state mass profiles that best fit that galaxy's Burkert profile which were rejected by the above criteria.  All of the values of $\Up$ above those computed are also rejected because they produce mass profiles even closer to the constant function.  In Table \ref{Upsilon_upbnd}, we have collected these upper bounds of $\Up$. In Figure \ref{badfitSextans}, for the galaxy Sextans, we present best fit static state mass profiles for the ground through fifth, tenth, and twentieth excited states for which $E(r_{h})$ is more than $80\%$ of the value of $E$ for the best fit constant function.

\begin{table}

    \begin{center}
        \begin{tabular}{c|cccccccc}
            Galaxy $\backslash$ State & 0 & 1 & 2 & 3 \\
            \hline & & & & \\
            Sextans & $\Up < 160$ & $\Up < 394$ & $\Up < 633$ & $\Up < 875$ \\
            Leo II & $\Up < 234$ & $\Up < 576$ & $\Up < 926$ & $\Up < 1279$ \\
            Fornax & $\Up < 35$ & $\Up < 87$ & $\Up < 139$ & $\Up < 192$ \\
            Leo I & $\Up < 57$ & $\Up < 141$ & $\Up < 226$ & $\Up < 312$ \\
            Sculptor & $\Up < 49$ & $\Up < 121$ & $\Up < 194$ & $\Up < 268$ \\
            Ursa Minor & $\Up < 82$ & $\Up < 202$ & $\Up < 325$ & $\Up < 449$ \\
            Carina & $\Up < 84$ & $\Up < 207$ & $\Up < 333$ & $\Up < 459$ \\
            Draco & $\Up < 45$ & $\Up < 111$ & $\Up < 179$ & $\Up < 246$
        \end{tabular}

        \vspace{.2 in}

        \begin{tabular}{c|cccc}
           Galaxy $\backslash$ State & 4 & 5 & 10 & 20 \\
           \hline & & & & \\
           Sextans & $\Up < 1116$ & $\Up < 1356$ & $\Up < 2460$ & $\Up < 4789$ \\
           Leo II & $\Up < 1632$ & $\Up < 1983$ & $\Up < 3597$ & $\Up < 7003$ \\
           Fornax & $\Up < 245$ & $\Up < 297$ & $\Up < 538$ & $\Up < 1048$ \\
           Leo I & $\Up < 398$ & $\Up < 484$ & $\Up < 877$ & $\Up < 1706$ \\
           Sculptor & $\Up < 342$ & $\Up < 416$ & $\Up < 754$ & $\Up < 1467$ \\
           Ursa Minor & $\Up < 572$ & $\Up < 695$ & $\Up < 1261$ & $\Up < 2455$ \\
           Carina & $\Up < 586$ & $\Up < 712$ & $\Up < 1292$ & $\Up < 2514$ \\
           Draco & $\Up < 314$ & $\Up < 382$ & $\Up < 692$ & $\Up < 1347$
        \end{tabular}
    \end{center}

    \caption{Upper bound values for $\Up$ corresponding to poor best fits of the Burkert mass profiles for each of the classic dwarf spheroidal galaxies.  The values in each column for each galaxy should be interpreted as an upper bound on the value of $\Up$, under the approximations explained in the paper, if that galaxy is best modeled by an $n^{\text{th}}$ excited state.  The units on $\Up$ are $\text{yr}^{-1}$.}

    \label{Upsilon_upbnd}

\end{table}

We observe from Table \ref{Upsilon_upbnd} that the upper bound values of $\Up$ increase as we increase the state we consider.  This is due to the following.  The rejected values of $\Up$ correspond to rejected hyperbolas of constant $\Up$, and hence constant $m r_{h}$.  Furthermore, the only qualitative difference between any two $n^{\text{th}}$ excited state mass profiles is the number of ripples in the initial increasing region of the profile.  For large $\Up$, the majority of a best fit $n^{\text{th}}$ excited state mass profile that is compared to the Burkert profile is the constant region which is shared by static state mass profiles for any $n$.  Thus the hyperbola corresponding to a rejected best fit ground state is close to the hyperbola corresponding to a rejected best fit $n^{\text{th}}$ excited state for any $n$.  In particular, there is a hyperbola of constant $m r_{h}$, for which the corresponding best fit $n^{\text{th}}$ excited state mass profiles for any $n$ are rejected by the above criteria.  Then, since $C_{mass}^{n}$ and $C_{radius}^{n}$ appear to monotonically increase as $n$ increases (see Table \ref{ScorrVal}), by equation (\ref{mass-hyperbola}), we would expect the same behavior for the value of $\Up$ in order for $m r_{h}$ to remain constant, which is what we observe in Table \ref{Upsilon_upbnd}.

Thus if a dwarf spheroidal galaxy is correctly modeled by a twentieth excited state or less, then an overall upper bound on the value of $\Up$ would be the upper bound corresponding to the twentieth excited state.  The least upper bound corresponding to the twentieth excited state over all eight galaxies is that value for the Fornax galaxy, which yields approximately that
\begin{equation}\label{Upupbnd}
  \Up < 1000 \text{ yr}^{-1}.
\end{equation}

\section{Utilized Approximations}

Now that we have presented our results, we list here the important approximations made in this paper which led to these results and explain briefly why we make them.
\begin{description}
  \item[Approximation 1:] Dark matter is correctly described by the wave dark matter model.
  \item[Approximation 2:] Dark matter halos around dwarf spheroidal galaxies are spherically symmetric.
  \item[Approximation 3:] The dwarf spheroidal galaxies used in this paper are in a state of dynamical equilibrium.
  \item[Approximation 4:] The spacetime metrics describing these dwarf spheroidal galaxies are static.
  \item[Approximation 5:] Wave dark matter predicts outcomes qualitatively similar to those of spherically symmetric static state solutions to the Einstein-Klein-Gordon equations.
  \item[Approximation 6:] The Burkert mass profiles computed by Salucci et al.\ \cite{Salucci11} fit the observational data very well.
  \item[Approximation 7:] The spacetime is in the low field limit, that is, $M \ll r$.
  \item[Approximation 8:] The spacetime is asymptotically Schwarzschild.
\end{description}

Approximation 1 is used because we are testing the wave dark matter model against observations.  Approximation 2 is a common approximation for dwarf spheroidal galaxies and is also necessary because we are comparing the wave dark matter model to the spherically symmetric Burkert mass profile.  Approximation 3 seems to be consistent with observations of dwarf spheroidal galaxies at least out to large radii \cites{Salucci11,Cote99}. Approximation 6 is reasonable given Figure~\ref{vel-disp}.  Approximations 7 and 8 are standard when modeling galaxies.

Approximations 4 and 5 are used to simplify the types of solutions to the Einstein-Klein-Gordon equations we consider.  We note here, however, that there is a question of the stability of the spherically symmetric static state solutions.  It is known that, if the corresponding total mass is not too large, the ground state is stable under perturbations \cite{Seidel90,Lai07} but that, on their own, the excited states are not \cite{Seidel98} regardless of their mass.  However, it has also been shown that a coupling of an excited state with a ground state can produce a stable configuration \cite{MSBS}.  We hypothesize that luminous matter distributions coupled with combinations of static states will produce a stabilizing effect allowing for more dynamically interesting systems to be physically plausible.

\section{Conclusions}

To summarize the results of this paper, we have drawn effectively two conclusions, which we list here.

\begin{conc}
  Given Approximations 1 through 8, a value of $\Up$ which yields one or more spherically symmetric static state mass profiles which match well the best fit Burkert mass profiles computed by Salucci et al.\ \cite{Salucci11} for each of the eight classical dwarf spheroidal galaxies is $$\Up = 50 \text{ yr}^{-1}.$$
\end{conc}

\begin{conc}
  Given Approximations 1 through 8 and Rejection Criteria \ref{rejcrit_1}, if the dark matter halos of all of the eight classical dwarf spheroidal galaxies are correctly modeled by $20^{\text{th}}$ excited states or less, then $$\Up < 1000 \text{ yr}^{-1}.$$
\end{conc}

For the interested reader, the Matlab code used for this paper to generate the spherically symmetric static states and to compute the best fits to a Burkert profile can be found on Bray's Wave Dark Matter Web Page at \url{http://www.math.duke.edu/\~bray/darkmatter/darkmatter.html}.

\section{Acknowledgements}

The authors gratefully acknowledge the support of National Science Foundation Grant \# DMS-1007063.

\bibliographystyle{amsalpha}
\bibliography{./References}

\end{document}